\documentclass[doublecol]{epl2} 

\usepackage{amssymb,amsfonts}
\usepackage{amsmath,hyperref,srcltx,cite}

\title{Temperatures in Grains and Plasma}

\author{Mario Liu}

\institute{Theoretische Physik, Universit\"{a}t T\"{u}bingen,\\ 72076 T\"{u}bingen, Germany, EC }

\abstract{
{Grains are widely assumed to be characterized by a single temperature -- derived either from the configurational entropy, or employing the kinetic theory. Yet granular media do have two temperatures, $T_g$ and $T$, pertaining to the grains and atoms. It is argued here that a two-temperature plasma yields a more useful analogy for grains than a molecular gas: (1)~Irreversible collisions also occur in plasma, to reach the equilibrium of equal temperature. (2)~The plasma energy is not linear in the two temperatures; it is quadratic in the temperature difference, minimal at equilibrium. Both points have valid analogues in grains, yielding useful insights.
}}

\begin{document}
\date{\today}\maketitle


\section{Introduction}
Two better-known notions of granular temperature  exist,  the kinetic temperature $T_k$ for granular gases, and the configurational temperature $T_{con}$ for grains at rest. Neither is conceptually fully satisfactory.

{The kinetic temperature $T_k$ is founded on the observation that granular gas in a shear flow displays velocity fluctuations $|\Delta v|$, similar to molecules in a gas. So $T_k$ is taken as the 
average kinetic energy per grain, $T_k\sim|\Delta v|^2$, same as with molecules. 
However, without a shear flow, a molecular gas is in equilibrium, and its temperature $T$ is 
thermodynamically defined. Enforcing a shear flow only increases this $T$ by viscous heating. Grains, on the 
other hand,  collide irreversibly and inelastically, and collapse onto a heap without a shear flow. Granular gas is therefore always off equilibrium; there is no equilibrium state to define a temperature. It is therefore not 
clear whether  $T_k$ has much in common with the conventional, thermodynamic temperature $T$, as 
keenly observed by Goldhirsch~\cite{Goldhirsch}. This has lead to the broad impression that grains are abnormal, since conventional thermodynamics is inapplicable. 

In this dilemma, and observing that grains,  being large, do not execute perceptible Brownian motion -- as if they were ``athermal" -- Edwards proposed a novel approach, to calculate a \textit{configurational  entropy} $s_{con}$, see eg.~\cite{kamrin}, as the logarithm of the many 
stable granular packings, or configurations. Initially, all configurations were assumed equally probable, though this conjecture turned out not to be generally 
true~\cite{PT}. What makes $s_{con}$ special, and intriguing, is the fact that all configurations have the same zero energy: Grains at 
rest possess no kinetic energy, and no deformation energy if infinite rigidity is assumed. Hence the usual temperature definition as a derivative of the energy
does not work, necessitating a novel definition, in which the volume replaces the energy. When grains do move, slowly, it is taken as an \textit{``exploration of the phase space''}, with the associated kinetic energy ignored.  

Such a $T_{con}$ has little in common with the thermodynamic temperature $T$, or the kinetic one $T_k$. For instance, it does not equilibrate with either, though this is a basic property of a temperature. And it is completely unclear how to bridge  $T_k$ and $T_{con}$, to form a more general temperature, valid for the states in between. Depending on how fast the grains move, the kinetic energy becomes impossible to ignore at some stage, long before the gaseous phase is reached. Then the basic  $s_{con}$-assumption of same zero energy for all configurations falls flat. 

A more fundamental objection against both $T_k$ and $T_{con}$ concerns the assumed analogy between grains and molecules. 
In calculating the entropy $s$, say,  of a molecular gas,  the internal degrees of freedom  (DoF) within the molecules -- such as the electronic and nucleonic ones --  may be neglected, because they possess energy gaps so large that they are,  in typical condensed matter settings,  quantum-mechanically frozen.  
Not so in grains, in which the microscopic DoF -- phononic and free-electronic ones -- are low lying. They actively interact with the grains, and this is why atoms collide elastically but grains undergo dissipative 
collisions, redistributing the energy from the granular DoF to  microscopic ones. 
Clearly, the equilibrium the system strives toward is given by maximal total entropy that includes all DoF, granular and microscopic. 

A general notion of temperature would certainly help to better understand granular media. It should hold for grains both at rest and fast moving, also for all granular states in between: gaseous, fluid and  solid. For this purpose, let us  
allow for the possibility that grains have, instead of one, two interacting temperatures, with two associated entropies: A conventional one $s$ that includes all microscopic DoF, and a granular one $s_g$ that includes the granular DoF. The temperatures are then thermodynamically defined as $T\equiv\partial w/\partial s$ and $T_g\equiv\partial w/\partial s_g$. 

A theory based on such a $T_g/T$-pair, called GSH, for \textit{granular solid hydrodynamics}, has been shown to be applicable to a wide range of granular experiments~\cite{Jiang2007c,granRexp}. 
However, the fundamental  question of the temperature definition remains,  casting doubts on GSH. 
It is whether such an equilibrium definition is permissible for a necessarily off-equilibrium system. Granular media are in equilibrium only as a static pile of sand, in which $T_g=T$ holds, but not if excited as fluid or gas, for $T_g\gg T$. 

We clarify this question by considering a seemingly far-fetched, yet surprisingly close analogue, the two-temperature  plasma, with the electronic and ionic temperature $T_e$ and $T_i$. This is a much more apt model for granular media than the molecular gas. For instance, same as in granular gas, irreversible collisions also occur in plasma, to reach the equilibrium $T_e=T_i$, without invalidating the definition of the temperatures.

A further analogy concerns the energy dependence of the temperatures. The energy $w$ of a molecular gas is linear in its temperature, $w\sim T$. Modeling on it, the kinetic temperature $T_k$ is also taken 
as proportional to the grains' energy. And because a granular gas has only kinetic energy, we have $T_k\sim |\Delta v|^2$.  In a plasma, perhaps surprisingly, the energy is not a sum of two terms, one $\sim T_e$, the other $\sim T_i$. The reason is, it has to have a 
minimum at $T_e=T_i$. More specifically, as we shall see below, the plasma energy is $\sim (T_e-T_i)^2$. 

The same reasoning holds for grains, their energy is also given as $\sim (T_g-T)^2$ -- and since $T_g\gg T$ almost always -- 
as $\sim T_g^2$.
Generally, both the kinetic and elastic energy of the grains need 
to be included in calculating $T_g$. In granular gases, however, collisions are rare, and there is little deformation of the grains. Then $T_g$ is, similar to $T_k$, given by the granular kinetic energy alone, or  $T_g\sim|\Delta v|\sim\sqrt{T_k}$. We conclude that although both $T_k$ and $T_g$ may be used to 
quantify  granular jiggling and velocity fluctuations $ |\Delta v|$, only $T_g$ is a bona fide temperature that equilibrates with other temperatures.
That is how $T_g$ connects to $T_k$. 

The connection to $T_{con}$ is rather more tedious.
A static sand pile is in equilibrium with equal temperatures,  $T_g=T$,  which is a result of maximal $\int{\rm d}^3r(s+s_g)$. In this context, $s_g\ll s$ is not relevant,  maximal $\int (s_g+s)\, d^3r$ is well approximated by maximal  $\int s\, d^3r$. 
The resultant Euler-Lagrange conditions are uniform $T$ and force balance among grains.  

The configurational entropy is an even smaller quantity, $s_{con}\ll s_g$. It includes only 
ways of stably packing the grains and excludes all moving configurations. Whether the consideration of $s_{con}$ yields additional information that goes beyond conventional thermodynamics, is not a question that can or should be answered here. It suffices for one to realize 
that understanding and adopting the plasma-grains analogy,  we find the behavior of grains to be utterly normal, and usefully amenable to the usual tools of  theoretical physics, including especially thermodynamics and statistical mechanics.

In the following section, we shall first consider a generic two-temperature system, to show that a thermodynamic  temperature definition remains valid, as long as the system is in  \textit{constrained and local equilibrium} (see the explanation below), even though it is not in complete and total equilibrium. Then we go on to consider the two-temperature plasma, and draw various useful analogies to the behavior of  two-temperature granular media. The sections ``plasma energy'' and ``granular heat'' are more quantitative. Here, it is shown that given a relaxation equation that accounts for the relaxation of the relevant temperature difference, one may infer the quadratic temperature dependence of the energy. The final section ``summary'' concludes the paper. }

\section{Two-temperature systems} The most elementary  two-temperature system consists simply of two vessels containing a gas of different  temperatures, $T_1$ and $T_2$. Both are each in equilibrium, with  $T_1, T_2$ thermodynamically well-defined. Bringing them into thermal contact, the total system is off-equilibrium during 
equilibration, yet $T_1, T_2$ remain well-defined if the contact is weak enough to not overly perturb the systems, such that the only change is in the temperatures.

The next system is a fully ionized plasma, with an electronic $T_e$ and an ionic $T_i$, both thermodynamically defined,  $T_e=\partial w/\partial s_e$,  $T_i=\partial w/\partial s_i$. 
The plasma is in equilibrium for $T_e=T_i$, though the two temperatures remained valid for $T_e\not= T_i$, because (due to the difference in the electronic and ionic masses) equilibration is  
slow. After any perturbations, the plasma thermalizes quickly, such that the electrons and ions are 
each in equilibrium. It then slowly relaxes $T_i-T_e$, the one off-equilibrium DoF, without  perturbing the two equilibria.

The situation, in which only a few macroscopic DoF are off equilibrium, occurs frequently. It is  referred to  as a \textit{constrained equilibrium}, and these DoF as  \textit{non-hydrodynamic}  variables, or \textit{slow, macroscopic} ones.  Including them, both thermodynamics and statistical mechanics work. 
Taking the free energy to depend on the order parameter $\psi$, a non-hydrodynamic variable, in the  Ginzburg-Landau theory to account for phase transitions epitomizes this approach.  

Another generalization of the equilibrium  concept is \textit{local equilibrium}, say when the temperature is non-uniform. Here, every volume element is in equilibrium, but neighboring elements are in slightly different ones.  

For time spans large compared to the collision time of grains, granular media are also in constrained and local equilibrium, in which the equilibrium definition of $T_g$ and $T$ holds. Note that the relaxation of $T_g-T$, being contingent on collisions, is slow on the scale of the collision time,
same as in plasma.



In a plasma, electrons and ions collide irreversibly for $T_e\not= T_i$. If $T_e>T_i$, the electrons move more quickly, dissipating their kinetic energy to heat up the ions.  The total energy of electrons and ions is, of course, conserved. 
In granular media, the grain-grain collisions are also irreversible, because atomic DoF are being heated up by the collisions. At the same time, the total energy of grains and the atomic DoF is conserved. The thermodynamic analogy between the two systems is therefore fairly complete, and inelastic granular collisions  do not invalidate the temperature definition. 

Nevertheless, there remains a basic obstacle to the plasma-grains analogy, which is: At  $T_e=T_i$, the plasma is in equilibrium. Its analogue in grains, the equilibrium at $T_g=T$, does not usually exist -- except in a static pile. 
Grains do have two types of DoF, the usual microscopic ones including electronic and phononic DoF,  and the much larger-scaled granular ones, pertaining to the velocity, rotation, position, deformation of grains. Yet due to the comparative massiveness of grains, we have $T_g\gg T$ for any granular motion, and grains melt or vaporize long before $T$ approaches $T_g$. 
On the other hand, this obstacle is based on a specific material property -- the melting temperature of the granular material -- not on a general principle. To complete our line of thoughts, let us therefore assume grains do maintain their integrity at any $T$, and never melt. What are then the ramifications?

Granular collisions are inelastic, and the restitution coefficient is  $r<1$ because the grains dissipate to heat themselves up.\footnote{The coefficient  $r$ is the quotient of final to initial 
velocity.} Same holds for the frictional forces between grains, sliding also heats up the 
grains. Both are a transfer of energy from $T_g$ to $T$, for $T_g>T$. 
Conversely, of course, we have $r>1$ for $T_g<T$, implying grains are being accelerated by  internal cooling. Similarly, if grains are initially at rest, $T_g=0$, Brownian motion is initiated. 

In equilibrium, at  $T_g=T$, collisions among grains are fully elastic, with $r=1$ and zero net transfer of energy. This is not different from a plasma at equilibrium, in which all collisions are elastic, or from Brownian particles jiggling forever in equilibrium, because their interaction with the ambient fluid is elastic.

Conceptually speaking, clearly, granular media are not necessarily off-equilibrium. A granular gas is, for $T=T_g$, in perfect equilibrium, and will persist forever, in a close vessel, without any external energy input.  
Having established that there is an conceptual equilibrium state, $T_g=T$, to define both temperatures, we conclude that a granular gas, liquid, or solid, at $T_g\gg T$, is in a constrained equilibrium, same as a plasma at $T_e\not=T_i$. 

It is useful to realize that this fact does have experimental consequences. For example, at the very instance one starts to shear static grains, it should be fully elastic, with vanishing plastic shear rates, because  $T_g=T$ initially.  This has already been observed frequently but interpreted as a small elastic region that each starting point in strain space possesses, within which motions

 are fully elastic.

A further analogy between plasma and grains that will prove useful concerns the energy. 
By applying the equipartition theorem, one typically takes the plasma energy as the sum of two terms, one $\sim T_e$, the other $\sim T_i$. However, this expression has no minimum at $\Delta T\equiv T_e-T_i=0$, and can only be valid for a plasma that, lacking any interaction between electrons and ions, fails to drift toward  $T_e=T_i$. There must be a term with a minimum at $\Delta T=0$. And as will be shown, it is of the form $\sim\Delta T^2=(T_e-T_i)^2$. For the same reason, the granular energy must also have a minimum at  $T_g=T$, and is again of the form $\sim(T_g-T)^2$. For $T_g\gg T$, this is hardly  different from $\sim T_g^2$.  

Both  $\Delta T$ and $T_g-T\approx T_g$ are difficult to measure. A usual thermometer needs to equilibrate with the host system, which works only in equilibrium, only for $T_g=T$ or $T_e=T_i$.  What we need is a thermometer that equilibrates exclusively with the subspecies, either the grains irrespective of $T$, or the electrons independent of the ions (and vice versa). This does not appear easy. 

One indirect solution is to measure the velocity fluctuations  $|\Delta v|$ of grains, assuming a given relation to $T_g$. This experiment has been performed successfully~\cite{benji}. 
As discussed above, we have $T_k \sim |\Delta v|^2$ in the kinetic theory that models on the molecular gas~\cite{haff,jenkins,savage,campbell, goldhirsch,luding2009}. The basis is  the equipartition theorem. And we have $T_g\sim|\Delta v|$ in GSH~\cite{Jiang2007c,Jiang2009, athermal,itai, stefan}, which is derived from thermodynamics that requires the energy being minimal at $T_g=T$. As will be argued below, 
the second is correct.

\section{Plasma energy} 
Given time, the temperatures of electrons and ions will equilibrate. Understanding how this happens and what the associated energy is, we draw useful information for grains. 
\footnote{The particle-particle interaction in plasma is long-ranged, not in grains.  Yet interactions are typically irrelevant for thermodynamic considerations. And as we shall see, they do not enter our discussion.}

If the plasma is sufficiently hot, the kinetic energy of  electrons and ions dominates. 
The average energy is typically taken as: $\frac32k_BT_e=\frac32m_e(|\Delta v|_e)^2$ per electron,  and $\frac32k_BT_i=\frac32m_i(|\Delta v|_i)^2$ per ion. ($m$ denotes the mass, and $k_B$ the Boltzmann constant.) 
The sum seems to favor the quadratic dependence, but this energy does not have a minimum at equilibrium $\Delta T\equiv T_e-T_i=0$. 
Denoting $T_e=T_{\rm eq}+\frac12\Delta T$, $T_i=T_{\rm eq}-\frac12\Delta T$ and inserting them into $w=\frac32k_B(n_eT_e+n_iT_i)$, with $n_e=n_i=n$ for a neutral plasma, we find
$w=3nk_BT_{\rm eq}$ does not depend on $\Delta T$ at all,  
yet it must. 

The theory of relaxation in a two-temperature plasma is developed by Landau~\cite{Landau1936,Landau1937} and Spitzer~\cite{spitzer}, and revisited more recently by Hazak et al.\cite{hazak}. Assuming weak coupling between electrons and ions, but a strong one among themselves (ie. uniform, yet unequal $T_e,T_i$), they obtain
\begin{align}\label{rel1}
\partial_tT_e\sim-{(T_e-T_i)}{(T_i/m_i+T_e/m_e)^{-3/2}}\,.
\end{align}
We  rewrite it, to linear order in $\Delta T$, as
\begin{align}
\label{rel2}
2\partial_t T_e=\partial_t \Delta T=-{\Delta T}/\tau,\quad \tau\sim {T_{\rm eq}^{3/2}},
\end{align}
identifying $\Delta T$ as the relaxing quantity. ($T_{\rm eq}$ remains constant if uniform, hence $2\partial_t T_e=\partial_t\Delta T$.)   
It is this equation that implies an energy contribution $\sim(\Delta T)^2$.

Consider a system with only one variable $A$, the energy density  $w(A)$ and the conjugate variable  $B\equiv\partial w/\partial A$. 
If the minimum is at $A=A_0$,  with $B=0$, the system is in equilibrium there. Expanding the energy in its vicinity, we have, to lowest order of $A-A_0$ and with $b(A)>0$,
\begin{align}\label{energy}
w-w_0=\frac{(A-A_0)^2}{2b}=\frac b2 B^2,\quad B=\frac{A-A_0}b.
\end{align}
Again expanding  $\partial_tA$ in $B$, noting both vanish in equilibrium, we obtain, again to lowest order and with $\alpha(A)>0$,  a relaxation equation, 
\begin{align}\label{rel3}
\partial_tA=-\alpha B=-(A-A_0)/\tau,\quad \tau=b/\alpha.
\end{align}
This excursion shows that relaxation and a parabolic energy contribution are generically related, and that Eq.(\ref{rel2}) is the result of a $(\Delta T)^2$-dependence of the energy. 
Hence, close to $\Delta T=0$, the plasma energy should read 
\begin{align}\label{kinE}
w=3nk_BT_{\rm eq}+\Delta T^2/2b.
\end{align}
As long as there is an electron-ion interaction, irrespective of its form, leading to a relaxation toward equilibrium, the second term must be included. 
It may be small (ie. $b$ large), and the relaxation slow (ie. $\tau\sim b$ large), as is the case for a weakly coupled plasma. Yet it is the only term that depends on $\Delta T$. (Interaction among electrons, or among ions, is not included explicitly, because we assume, as do Eqs.(1,2), that it is sufficiently strong to maintain uniformity of $T_e$ and $T_i$ at all time.)

The conclusion to draw here is that \textit{the equipartition theorem} holds for $T_{\rm eq}$,  
but not for $\Delta T$.  According to this theorem, every DoF of a  classical, conservative system that enters the Hamiltonian quadratically, yields the energy contribution of $\frac12k_BT$. It holds for  $T_{\rm eq}$, cf. Eq.(\ref{kinE}), because the plasma as such is conservative. It does not hold for the relaxing variable $\Delta T$, because electrons and ions  are not separately conservative for $\Delta T\not=0$. They are only for  $\Delta T=0$, or if their interaction is negligible, for a time span much smaller than the relaxation time.

{Even though the second term of Eq.(\ref{kinE}) holds only for $\Delta T\ll T_{\rm eq}$, as terms of higher order in $\Delta T$ become important otherwise, the conclusion that the equipartion theorem only applies to $T_{eq}$, and does not apply to $\Delta T$, $T_e$ or $T_i$ is more general, implying that the energy is neither proportional to $T_e$, nor to $T_i$. It remains especially valid for $T_e\ll T_i$, or $T_e\gg T_i$. }

\section{Granular heat}\label{sec4}
The plasma results enable us to draw useful inferences for grains, by employing the analogy
\begin{align}\label{eq6}
T_{\rm eq}\to T, \quad \Delta T\to T_g-T\approx T_g.
\end{align}
The first  expression holds because there are many more DoF represented by $T$ than by $T_g$ -- in contrast to a neutral plasma,  in which the number of DoF is the same for both populations, hence  $2T_{\rm eq}=T_e+T_i$ holds.

Equilibrium, $T_g=T$, reigns when the grains are at rest, say in a sand pile, in which the Brownian motion is imperceptible. It also holds when grains are flowing very, very slowly, and jiggling slightly more strongly, which heats up the grains, until both temperatures are equal -- of course as long as the grains remain intact.~\footnote{
{One may question whether very slow shear rates are at all possible, as granular media typically move through discrete reorganizational events. However, the slow shear rates accounted for by a macroscopic, constitutive model may indeed be such events that happen rarely --  the slower the rates, the less frequent the occurrences.}} 
Interaction among grains is then necessarily elastic, with a restitution coefficient of one, and zero plastic rates. This is probably related to the prediction of GSH  (made from different considerations) for vanishingly small shear rates, 
at which the plastic rate diminishes, and granular media become  \textit{quasi-elastic}~\cite{granR4,granRexp}.

Next, we consider the explicit form of granular heat $w_T$ -- the grain-size part of the kinetic and potential energy. The kinetic energy per grain is given by 
$\frac32m|\Delta v|^2$. As the grains are sufficiently large, one probably needs to include the rotational DoF as well, doubling the energy if both types of DoF equilibrate well. 
For a close-packed granular ensemble, the compressed grains form a network of linear oscillators, with six DoF per grain, three kinetic and  three potential.
Then again $w_T=3mn|\Delta v|^2$, since the kinetic and potential contributions are equal. If the density is somewhat less, ``Rattlers"  exist that are not members of the network. But they do have kinetic and some rotational contributions. 
Hence granular heat is,
\begin{align}\label{eq8}
w_T=\alpha (3/2) nm|\Delta v|^2,\quad \alpha\leq 2.
\end{align}

The kinetic theory delivers two results from this that are contradictory.
First, applying the equipartition theorem, attributing  the energy contribution  $k_BT_k/2$ for every DoF, it takes granular heat as $w_T=\alpha(3/2)nk_BT_k$, or $T_k\sim|\Delta v|^2$. 
Second, $T_k$ is shown to obey a relaxation equation that, neglecting all other terms, reads~\cite{haff,jenkins,savage,campbell,goldhirsch,luding2009}
\begin{align}\label{eq9}
\partial_tT_k=-T_k/\tau+\cdots,\quad 1/\tau\sim\sqrt{T_k}.
\end{align}
Such an equation, as we concluded from considering plasma, implies an energy dependence $w_T\sim (T_k-T)^2$.

Faced with the dilemma of a quadratic versus linear dependence on $T_k$,
there are two reasons to stick with the relaxation equation (\ref{eq9}), and choose the quadratic one. First, it reflects the general phenomenon of equilibration; second,  the equipartition theorem does not hold for $T_k$, as it is -- same as 
{$T_e$ for $T_e\gg T_i$ -- a dissipative, relaxing variable, accounting for the energy transfer  from the grains to the microscopic DoF.}  Furthermore, taking the energy as $\sim T_g^2$ leaves all results of the kinetic theory intact, if one identifies the granular temperature as $T_g\sim \sqrt{T_k}$.

This is, in fact, the granular heat expression of  GSH, employed without realizing the intricacies discussed here,
\begin{align}\label{eq11}
w_T=(T_g-T)^2/2b\approx T_g^2/2b.
\end{align}
Equating this to Eq.(\ref{eq8}), we find $T_g\sim|\Delta v|\sim\sqrt{T_k}$.
The relaxation of $T_g$ is (neglecting other terms) given as
\begin{align}\label{eq12}
\partial_tT_g=-T_g/\tau+\cdots,\quad 1/\tau=\alpha_0+\alpha_1{T_g}.
\end{align}
The quasi-elastic regime mentioned above is given by $\alpha_0\gg\alpha_1{T_g}$,  and the rate-independent regime by  $\alpha_0\ll\alpha_1{T_g}$, which is the same as Eq.(\ref{eq9}), see\cite{granR4,granRexp}.  

In Eq.(\ref{kinE}) of plasma, the term $\sim(\Delta T)^2$ was seen as the lowest order one, with higher order terms becoming relevant as $\Delta T$ grows. This is different here:  Any terms of higher order in $T_g$ would contradict Eq.(\ref{eq9}). As long as it holds, $T_g^2/2b$ is the only term we may include, and  $T_g\sim|\Delta v|$ holds in granular solid, liquid and gas, though the coefficient changes with the density. 

In the limit of rarefied granular gases, $n\to0$ (implying $b,\tau\to\infty$), since $T_g$ relaxes only during collisions that rarely happen, the grains may be taken as elastic for the time span $t\ll\tau$, and $T_k\sim|\Delta v|^2$ holds.



\section{Summary} The two-temperature plasma is an apt model for granular media: Equilibrium reigns when the two  temperatures are the same, turning into a constrained one if they are not. Dissipative collisions are what cause the temperature difference to relax toward equilibrium. 

Contrast this with taking $T_k$ as the sole temperature, as the analogue of $T$ in a molecular gas. Since the difference between  inelastical grains and elastic molecules is hard to bridge, grains appear abnormal, thermodynamics and statistical mechanics seem inapplicable. Embracing two temperatures, the grains are normal again, and usefully amenable to the usual tools of  theoretical physics.

\acknowledgements 
\noindent I am grateful to Eran Bouchbinder of Technion, who clarified for me, in numerous emails, the reigning understanding of  temperatures in granular media.

\end{document}